\documentclass[12pt]{article}
\usepackage{graphicx}
\usepackage{amsmath, amssymb}
\usepackage{array} 
\usepackage{tikz}
\usepackage{tikz-cd}
\usepackage{bussproofs}
\usepackage{booktabs}
\usepackage{geometry}
\usepackage{mathtools}
\usepackage{hyperref}
\usepackage{titlesec}
\usepackage{hyperref}
\usepackage{enumitem}    
\usepackage{tabularx}
\usepackage{tikz-qtree}

\usepackage{caption}
\usepackage{mathtools}                %
\usepackage{bm}                       
\usepackage{mathrsfs}      
\usepackage{float}
\usepackage{booktabs}

\usepackage{subfigure}
\usepackage{subcaption}

\newtheorem{theorem}{Theorem}[section]

\newtheorem{definition}[theorem]{Definition}

\title{Structural Origin and the Minimal Syntax of NP-Hardness:
Analysis of SAT from Syntactic Generativity  and Compositional Collapse}
\author{Yumiko Nishiyama}
\date{October 2025}

\begin{document}

\maketitle

\begin{abstract}
This paper proposes a structural reinterpretation of NP-hardness within the framework of Construction Defining Functionality (CDF), which models computation as layered transformations between syntax (\(S_{\mathrm{syn}}\)), semantics (\(S_{\mathrm{sem}}\)), and logic (\(S_{\mathrm{log}}\)). We argue that the intractability of NP-complete problems, exemplified by SAT, is not a result of algorithmic lack but of structural breakdown—specifically, the failure of compositional mappings across layers.

By contrasting 2SAT and 3SAT, we identify the critical role of clause-level syntax in determining semantic explosiveness. We formally show that for \(k \geq 3\), the map \(\alpha_f: S_{\mathrm{syn}} \rightarrow S_{\mathrm{sem}}\) becomes exponentially expansive, while the compositional identity \(\gamma_f = \beta_f \circ \alpha_f\) collapses. In contrast, 2SAT preserves compositionality via implicative, linearly structured syntax.

From these observations, we derive a broader claim: the syntax of SAT represents a generative minimum for NP-hardness. It is the minimal structural class sufficient to encode the semantic and logical complexity of any NP problem. Consequently, attempts to "simplify" NP-complete problems into polynomial-time solvable forms require altering their syntactic generators—thus transforming the identity of the problem itself.

We conclude that NP-hardness emerges from syntactic generativity and compositional collapse, positioning SAT not merely as a representative of NP, but as the structural origin of its computational hardness.
\end{abstract}

\section{Introduction and the Basic Concept of CDF}

The classical notion of a “function” in mathematics has long been conflated with its algorithmic counterpart in computation.  
In both notation and practice, functions such as
\[
f(x) = x + 1
\]
have been treated as if they simultaneously represent a symbolic rule, a mapping of meanings, and a computational procedure.

This conflation is understandable—but structurally misleading.

\subsection*{From Informal Functionality to Structural Clarity}

Historically, functions and algorithms were treated as interchangeable due to their shared operational behavior.  
However, from a structural standpoint, being a “function” involves far more than defining outputs from inputs.  
It requires preserving a layered correspondence between:

\begin{itemize}
    \item a \textbf{syntactic generation process},
    \item a corresponding \textbf{semantic interpretation},
    \item and a coherent \textbf{logical execution}.
\end{itemize}

These three layers—syntax, semantics, and logic—are often assumed to be aligned, but in complex systems (e.g., 3SAT), they can become misaligned, leading to combinatorial explosion and logical intractability.

To make these layers explicit, we introduce the \textbf{Construction Defining Functionality (CDF)} framework.

\vspace{1em}

\subsection*{The CDF Perspective on Functionhood}

The function notation above, though deceptively simple, already contains these three intertwined components:

\begin{itemize}
    \item The expression \( x + 1 \) on the right-hand side represents a \textbf{syntactic generative operation}; iterating this gives rise to a symbolic pattern—what we call the \emph{syntactic space}.
    \item The notation \( f(x) \) abstracts over this operation to define input-output mappings—this forms the \emph{semantic space}.
    \item The equality sign \( = \) enforces logical correspondence between symbol and meaning—this constitutes the \emph{logical space}.
\end{itemize}

Thus, what we casually call a "function" actually encodes a deep structural coordination among three representational levels.  
The CDF framework formalizes this structure.

\subsection*{Definition: Generated Structural Space via CDF}

Given a function \( f : X \to Y \), we define its CDF structure as a triple of spaces:

\begin{itemize}
    \item \( S_{\mathrm{syn}}(f) \): the \emph{syntactic component}, describing the symbolic form and generative rules of the function,
    \item \( S_{\mathrm{sem}}(f) \): the \emph{semantic component}, defining the space of input-output relations or meanings,
    \item \( S_{\mathrm{log}}(f) \): the \emph{logical component}, encoding inference procedures, computation traces, or proof trees.
\end{itemize}

These are connected by the following intercomponent mappings:

\[
\begin{tikzcd}
S_{\mathrm{syn}}(f) \arrow[r, "\alpha_f"] \arrow[dr, "\gamma_f"'] & S_{\mathrm{sem}}(f) \arrow[d, "\beta_f"] \\
& S_{\mathrm{log}}(f)
\end{tikzcd}
\]

\begin{itemize}
    \item \(\alpha_f : S_{\mathrm{syn}} \to S_{\mathrm{sem}}\): maps symbolic operations to interpreted relations.
    \item \(\beta_f : S_{\mathrm{sem}} \to S_{\mathrm{log}}\): maps semantic relations to logical derivations.
    \item \(\gamma_f : S_{\mathrm{syn}} \to S_{\mathrm{log}}\): composes the above two or defines direct procedural interpretation.
\end{itemize}

\subsection*{Concrete Example: \( f(x) = x + 1 \)}

\begin{itemize}
    \item \( S_{\mathrm{syn}}(f) \): syntactic expressions like \( f(0), f(1), f(2), \ldots \) built via \( x + 1 \),
    \item \( \alpha_f \): maps \( f(3) \) to the pair \( (3, 4) \),
    \item \( \beta_f \): interprets \( (3, 4) \) as a step in a computation or derivation \( 3 \vdash 4 \),
    \item \( \gamma_f \): maps \( f(3) \) directly to a derivation showing that 4 is the result of applying \( +1 \) to 3.
\end{itemize}

In this case, all mappings are well-behaved: the system is compositional, the layers align, and the function behaves as expected.

\subsection*{From Tractable Functions to Intractable Systems}

The power of CDF lies in its ability to distinguish when this compositional harmony breaks down.

- In \textbf{2SAT}, the CDF mappings remain stable: clauses can be transformed into implications; inference proceeds linearly.
- In \textbf{3SAT}, the \(\alpha_f\) map becomes \emph{explosive}: compact clauses generate a combinatorially vast semantic space, destroying the clean transition from syntax to logic.

This breakdown explains the root of NP-hardness—not merely as a failure of algorithm design, but as a structural collapse in the identity of a function.

\paragraph{CDF as a Criterion for Functionhood}

Thus, we propose:  
\emph{What makes something function-like is not just having inputs and outputs, but preserving a stable, interpretable, and derivable mapping between syntactic symbols and logical consequences.}

CDF provides a formal tool for assessing this structural integrity.

\section{Formal Definition of CDF Structures}

\subsection{Basic Components of Construction Defining Functionality (CDF)}
\begin{definition}[Generated Structural Space via CDF]

Given a function \( f : X \to Y \), the \emph{structural space} \( S(f) \) generated by \( f \) is defined as the family of sets comprising its syntactic iterative application, semantic syntactic expansion, and logical typing. The properties of \( S(f) \) are collectively called the \textbf{Construction Defining Functionality (CDF)} of \( f \).
\end{definition}

\begin{definition}[Structural Property Sets of CDF]

Let \( f : X \to Y \) be a function. The \emph{Construction Defining Functionality (CDF)} of \( f \) is defined as the triple of structural property sets:
\[
S(f) = \left( S_{\mathrm{syn}}(f),\ S_{\mathrm{sem}}(f),\ S_{\mathrm{log}}(f) \right),
\]
where
\begin{itemize}
    \item \( S_{\mathrm{syn}}(f) \) is the \emph{syntactic property set}, describing symbolic patterns and generative structure of \( f \)'s operation.
    \item \( S_{\mathrm{sem}}(f) \) is the \emph{semantic property set}, encoding input-output relations and interpretations.
    \item \( S_{\mathrm{log}}(f) \) is the \emph{logical property set}, reflecting computational or proof-theoretic realizations of \( f \).
\end{itemize}
\end{definition}

\begin{definition}[Intercomponent mappings]

Given the structural space S(f),
we define the \emph{intercomponent mappings}

such that:

\begin{itemize}
    \item \( \alpha_f: S_{\text{syn}} \rightarrow S_{\text{sem}} \) captures the unfolding of structure via iterated application.
    \item \( \beta_f: S_{\text{sem}} \rightarrow S_{\text{log}} \) captures interpretation, evaluation, or logical resolution.
    \item \( \gamma_f: S_{\text{syn}} \rightarrow S_{\text{log}} \) is the direct mapping representing \( f \).
\end{itemize}
\end{definition}

\subsection{CDF-Completeness and Stability/Explosiveness Definitions}

\begin{definition}[CDF-Completeness]
Let \( f : X \to Y \) be a function. We say that \( f \) is \textbf{CDF-complete} if the following three conditions are satisfied:
\begin{enumerate}
    \item The structural components \( S_{\text{syn}}(f), S_{\text{sem}}(f), \) and \( S_{\text{log}}(f) \) are all disjunctively definable, meaning they can be constructed from symbolic or logical primitives via rule-based or case-based generation.
    \item The intercomponent mappings \( \alpha_f : S_{\text{syn}} \to S_{\text{sem}}, \ \beta_f : S_{\text{sem}} \to S_{\text{log}}, \ \text{and} \ \gamma_f : S_{\text{syn}} \to S_{\text{log}} \) are in a state of structural balance, i.e., no single layer dominates or collapses the structure of the others, and the layers maintain mutual coherence.
    \item All three intercomponent mappings are stable, as defined below.
\end{enumerate}
\end{definition}

\begin{definition}[Stability, Explosiveness]
Let \( \varphi : A \to B \) be a (possibly partial) mapping between structural spaces, where both \( A \) and \( B \) are equipped with a size or complexity measure

\[
|\cdot| : A \cup B \to \mathbb{R}_{\geq 0}.
\]

\begin{itemize}
    \item \textbf{Stability}: The mapping \( \varphi \) is stable if there exist constants \( C > 0 \) and \( k \in \mathbb{N} \) such that for all \( a \in \mathrm{dom}(\varphi) \),
    \[
    |\varphi(a)| \leq C \cdot |a|^k.
    \]
    \item \textbf{Explosiveness}: The mapping \( \varphi \) is explosive if there exists a polynomially bounded sequence \( (a_n) \subset A \) such that
    \[
    \limsup_{n \to \infty} \frac{\log |\varphi(a_n)|}{\log n} = +\infty.
    \]
\end{itemize}
\end{definition}

\subsection{Compositional Coherence and Semantic Stability}

\begin{definition}[Compositional Coherence]
A CDF diagram is said to be \emph{compositionally coherent} if the following equality holds:

\[
\gamma_f = \beta_f \circ \alpha_f
\]

In this case, the function \( f \) is said to be \textbf{structurally coherent}, meaning that its logical behavior is fully mediated by its semantic structure.
\end{definition}

\noindent This coherence allows for modular analysis and controlled complexity propagation.

\begin{definition}[Semantic Stability]
The mapping \( \alpha_f \) is said to be \textbf{semantically stable} if the semantic space \( S_{\text{sem}}(f) \) grows polynomially (or sub-exponentially) with respect to the input size.

Conversely, \( \alpha_f \) is said to be \textbf{explosive} if \( S_{\text{sem}}(f) \) grows exponentially or worse.
\end{definition}

\subsection{Classification of CDF Structures}

Based on the behavior of \( \alpha_f, \beta_f, \gamma_f \), we define the following classes:

\begin{itemize}
    \item \texttt{ComCDF} (Compositionally Coherent CDF): \( \gamma_f = \beta_f \circ \alpha_f \), with stable \( \alpha_f \).
    \item \texttt{ExpCDF} (Explosive CDF): \( \gamma_f \ne \beta_f \circ \alpha_f \), or \( \alpha_f \) is unstable.
    \item \texttt{Semi-ExpCDF}: Intermediate case: \( \alpha_f \) partially unstable but diagram retains partial coherence.
\end{itemize}

\section{Syntactic and Semantic Approaches to Logical Proof}

Let us consider the formula:
\[
f(x) = A \rightarrow (B \rightarrow A)
\]
This formula is a tautology in propositional logic. We analyze it both semantically and syntactically.

\subsection*{Semantic Evaluation}

With two propositional variables $A$ and $B$, the truth table has $2^2 = 4$ combinations:

\begin{center}
\begin{tabular}{cccc}
\toprule
$A$ & $B$ & $B \rightarrow A$ & $A \rightarrow (B \rightarrow A)$ \\
\midrule
T & T & T & T \\
T & F & T & T \\
F & T & F & T \\
F & F & T & T \\
\bottomrule
\end{tabular}
\end{center}

The formula evaluates to true under all interpretations, confirming its semantic validity.

\subsection*{Syntactic Derivation (Natural Deduction)}

Using a natural deduction system, the formula can be proved as follows:

\begin{enumerate}
    \item Assume $A$ \hfill (Assumption)
    \item Assume $B$ \hfill (Assumption)
    \item \quad\quad $A$ \hfill (from 1)
    \item \quad $B \rightarrow A$ \hfill ($\rightarrow$-introduction: 2--3)
    \item $A \rightarrow (B \rightarrow A)$ \hfill ($\rightarrow$-introduction: 1--4)
\end{enumerate}

\quad Thus, $A \rightarrow (B \rightarrow A)$ is syntactically derivable using inference rules.

\subsection*{Semantic Explosion: Growth of the Truth Value Space}

One critical difference between syntactic and semantic approaches is the behavior of the semantic space under increasing logical complexity.

Let $n$ be the number of propositional variables. Then:
\[
|S_{\mathrm{sem}}(f)| = 2^n
\]
This exponential growth leads to what we term \textbf{semantic explosion}: the rapid expansion of the truth value space, making semantic evaluation intractable as $n$ increases.

\begin{center}
\begin{tabular}{cc}
\toprule
Number of Variables ($n$) & Truth Assignments ($2^n$) \\
\midrule
1 & 2 \\
2 & 4 \\
3 & 8 \\
4 & 16 \\
5 & 32 \\
6 & 64 \\
\bottomrule
\end{tabular}
\end{center}

While semantic evaluation provides direct insight into the meaning of formulas, its cost grows exponentially. In contrast, syntactic proof—though sometimes cumbersome—remains structurally bounded and manageable.

\subsection*{Comparative Observations}

\begin{center}
\begin{tabular}{p{4cm}p{5.5cm}p{5.5cm}}
\toprule
\textbf{Aspect} & \textbf{Semantic Evaluation} & \textbf{Syntactic Derivation} \\
\midrule
Method & Evaluate all truth assignments & Apply formal inference rules \\
Growth & Exponential in variables & Depends on formula complexity \\
Clarity & Intuitive meaning & Symbolic and formal \\
Drawbacks & Combinatorial explosion & Tedious or non-intuitive steps \\
CDF Mapping & $\alpha_f: S_{\mathrm{syn}} \to S_{\mathrm{sem}}$ & $\gamma_f: S_{\mathrm{syn}} \to S_{\mathrm{log}}$ \\
\bottomrule
\end{tabular}
\end{center}

This comparison shows how the CDF framework allows us to differentiate and analyze both the strengths and limitations of syntax-based and semantics-based reasoning.

\section{Reconsidering the Relationship between SAT and NP}

The conventional understanding of the relationship between SAT and NP is fundamentally reductionist and external:

\begin{itemize}
    \item The SAT problem is considered the \textbf{target of reduction} for all NP problems.
    \item Any decision problem in NP can be reduced to SAT in polynomial time (Cook–Levin Theorem).
    \item SAT functions as the \emph{representative} problem of NP, used to define NP-completeness.
\end{itemize}

While analytically powerful, this view is \textbf{external and operational}: it focuses on algorithmic reducibility and treats SAT as a convenient representative for classification purposes.

\subsection*{From Representative to Generator: A CDF Perspective}

Using the CDF framework, we may reframe SAT not merely as the endpoint of reductions but as a \textbf{structural generator} of NP-complexity.

\begin{quote}
SAT embodies syntactic constructions whose internal structure inherently gives rise to the semantic and logical complexity that defines NP.
\end{quote}

In terms of the CDF triad:

\begin{itemize}
    \item \textbf{Syntactic layer ($S_{\mathrm{syn}}$)}: The conjunctive normal form (CNF) used in SAT consists of disjunctions of literals, globally combined via conjunctions. This leads to non-local interactions and high combinatorial potential.
    \item \textbf{Semantic layer ($S_{\mathrm{sem}}$)}: The evaluation of satisfiability across $2^n$ truth assignments creates a vast, non-constructively distributed solution space.
    \item \textbf{Logical layer ($S_{\mathrm{log}}$)}: The act of finding a satisfying assignment involves implicit proof search, non-deterministic branching, and backtracking — reflecting the computational hardness of NP.
\end{itemize}

\subsection*{Semantic Explosion and Logical Non-Constructivity}

SAT is not difficult merely because we do not (yet) know efficient algorithms for it. Rather, its \emph{syntactic structure} naturally \emph{generates} hard computational behavior:

\begin{itemize}
    \item Each clause restricts part of the truth space but does not localize solutions.
    \item Conjunctions of clauses intersect in a non-trivial way, leading to semantic explosion in $S_{\mathrm{sem}}$.
    \item The structure in CNF causes logical exploration to require non-determinism and backtracking (complex proof trees in $S_{\mathrm{log}}$).
\end{itemize}

\subsection*{SAT as an Intrinsic Source of NP Complexity}

Thus, in the CDF view:

\begin{quote}
SAT is not merely a \textbf{representative} of NP, but an \textbf{intrinsic generator} of NP-complexity. Its syntactic design induces the semantic and logical properties that define the class NP.
\end{quote}

This structuralist viewpoint complements the classical complexity-theoretic interpretation, and offers deeper insight into why NP-complete problems are hard: they internalize the combinatorial and logical constraints that give rise to computational intractability.

\section{Concrete Instantiations: 2SAT vs 3SAT through CDF}

To make the distinction between structural simplicity and complexity concrete, we now apply the CDF framework to two canonical cases: 2SAT and 3SAT.

Although both are subsets of SAT, they exhibit drastically different behavior in their semantic and logical layers, rooted in their syntactic differences.

This section demonstrates how minor changes in the syntactic structure (e.g., moving from two to three literals per clause) lead to:

\begin{itemize}
    \item a shift from local to non-local semantic behavior,
    \item the breakdown of compositional mappings in CDF,
    \item and a qualitative leap in logical complexity, such as the necessity of backtracking.
\end{itemize}

From the CDF perspective, if each clause is reducible to an implication and the resulting structure forms a linear, tree-like dependency (as in 2SAT), the logical inference proceeds sequentially like:

\[
\texttt{•} \Rightarrow \texttt{•} \Rightarrow \texttt{•}
\]

This allows efficient, deterministic reasoning in polynomial time. In contrast, when the internal mappings (such as \( \alpha_f \), \( \beta_f \), and \( \gamma_f \)) become unbalanced—as in 3SAT—the logic structure fragments, leading to branching, backtracking, and exponential growth in the solution space. This structural imbalance is a key generator of computational intractability.

\subsection{2SAT Example: \( f_{\text{2SAT}} = (\lnot x \lor y) \land (\lnot y \lor z) \)}

\subsection*{Syntactic Component \( S_{\text{syn}}(f) \)}

\[
S_{\text{syn}}(f) = \{ (\lnot x \lor y), (\lnot y \lor z) \}
\]

\subsection*{Semantic Component \( S_{\text{sem}}(f) \)}

Each clause can be translated into an implication:

\[
(\lnot x \lor y) \iff (x \rightarrow y), \quad (\lnot y \lor z) \iff (y \rightarrow z)
\]

Thus:

\[
S_{\text{sem}}(f) = \{ x \rightarrow y,\ y \rightarrow z \}
\]

\subsection*{Logical Component \( S_{\text{log}}(f) \)}

The solver can propagate assignments using the implication graph:

\[
x = \text{true} \Rightarrow y = \text{true} \Rightarrow z = \text{true}
\]

This leads to a linear, deterministic derivation tree. No backtracking is required.

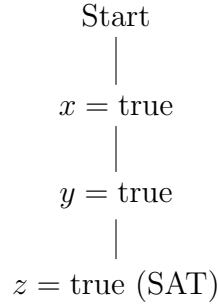
\begin{center}
\begin{tikzpicture}[sibling distance=2.5cm, level distance=1.2cm]
\node {Start}
  child { node {$x=\text{true}$}
    child { node {$y=\text{true}$}
      child { node {$z=\text{true}$ (SAT)} }
    }
  };
\end{tikzpicture}
\end{center}

\subsection*{Commutativity Check}

\begin{align*}
\alpha_f &: (\lnot x \lor y) \mapsto x \rightarrow y \\
\beta_f &: x \rightarrow y \mapsto \text{logical step: } x \vdash y \\
\gamma_f &: (\lnot x \lor y) \mapsto \text{direct solver inference}
\end{align*}

\[
\gamma_f = \beta_f \circ \alpha_f \quad \text{(Compositionality holds)}
\]

\subsection{3SAT Example: \( f_{\text{3SAT}} = (\lnot x \lor \lnot y \lor z) \)}

\subsection*{Syntactic Component \( S_{\text{syn}}(f) \)}

\[
S_{\text{syn}}(f) = \{ (\lnot x \lor \lnot y \lor z) \}
\]

\subsection*{Semantic Component \( S_{\text{sem}}(f) \)}

This clause is not reducible to a simple implication. It evaluates to false only under one assignment:

\[
x = \text{true},\ y = \text{true},\ z = \text{false}
\]

The satisfying assignments form a non-local, fragmented region in the solution space.

\subsection*{Logical Component \( S_{\text{log}}(f) \)}

The solver must explore assignments via backtracking to avoid the single falsifying case:

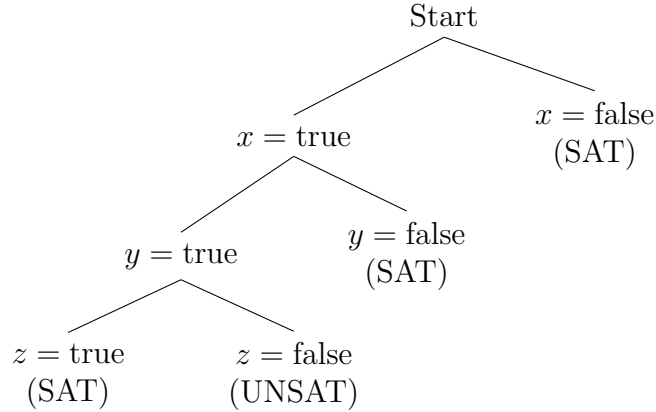
\begin{center}
\begin{tikzpicture}[
  level distance=1.6cm,
  level 1/.style={sibling distance=4cm},
  level 2/.style={sibling distance=3cm},
  every node/.style={align=center}
]
\node {Start}
  child { node {$x = \text{true}$}
    child { node {$y = \text{true}$}
      child { node {$z = \text{true}$\\(SAT)} }
      child { node {$z = \text{false}$\\(UNSAT)} }
    }
    child { node {$y = \text{false}$\\(SAT)} }
  }
  child { node {$x = \text{false}$\\(SAT)} };
\end{tikzpicture}
\end{center}

\subsection*{Commutativity Check}

\[
\begin{array}{rcl}
\alpha_f & : & (\lnot x \lor \lnot y \lor z) \mapsto \text{non-implicative constraint} \\
\beta_f & : & \text{semantics} \mapsto \text{non-deterministic solver behavior} \\
\gamma_f & : & \text{clause} \mapsto \text{backtracking search tree}
\end{array}
\]

\[
\gamma_f \neq \beta_f \circ \alpha_f \quad \text{(Compositionality fails)}
\]

\subsection*{Summary}

This comparison illustrates that the logical layer (\( S_{\text{log}} \)) of SAT problems is highly sensitive to small syntactic changes.

While 2SAT remains structurally tame due to compositional propagation, 3SAT introduces complexity via branching and non-determinism, confirming its role as an intrinsic generator of NP-hardness in the CDF framework.

\begin{center}
\begin{tabular}{|l|c|c|}
\hline
 & \textbf{2SAT} & \textbf{3SAT} \\
\hline
Derivation Tree Size & Small (Linear) & Larger (Branching) \\
Derivation Depth & Shallow (Deterministic) & Variable (Depends on conflicts) \\
Backtracking Required & No & Yes \\
Search Strategy & Unit propagation & Non-deterministic branching \\
Logical Mapping \( \gamma_f \) & Simple, Compositional & Complex, Non-compositional \\
\hline
\end{tabular}
\end{center}

\section{Generalization to \( k \)-SAT and Growth Theorem of \( \alpha_f \)}

Building upon the examples of 2SAT and 3SAT, we now consider the general case of \( k \)-SAT for \( k \geq 3 \).

\subsection*{Growth Theorem of \( \alpha_f \)}

\textbf{Theorem:} For \( k \geq 3 \), the syntactic unfolding \( \alpha_f \) exhibits exponential growth with respect to the input size.

\[
|\alpha_f| \sim O(c^n), \quad c > 1
\]

where \( n \) is the number of variables or clauses.

This contrasts sharply with the case \( k=2 \), where \( \alpha_f \) grows only polynomially or linearly.

\subsection*{Classification in the CDF Framework}

\begin{itemize}
  \item For \( k=2 \), the formula structure admits \textbf{ComCDF} behavior:
    \[
    \gamma_f = \beta_f \circ \alpha_f
    \]
    implying compositionality and efficient deterministic solving.

  \item For \( k \geq 3 \), the formula exhibits \textbf{ExpCDF} characteristics:
    \[
    \gamma_f \neq \beta_f \circ \alpha_f
    \]
    reflecting exponential syntactic explosion and non-compositional complexity requiring backtracking and non-determinism.
\end{itemize}

This theorem formally captures the qualitative jump in complexity between 2SAT and 3SAT, aligning with their known computational complexity classes (2SAT in P, 3SAT NP-complete).

\subsection*{Implications}

- The exponential growth in \( \alpha_f \) syntactic unfolding explains the combinatorial explosion in the semantic solution space and logical proof search space.

- This reinforces the view of SAT as a \textit{generator} of computational hardness within NP, consistent with the CDF perspective.

- The classification into ComCDF and ExpCDF provides a structural, formal framework for understanding the intrinsic difficulty of different SAT variants.

\section{Exponential Lower Bound of $\alpha_f$ in $k$-SAT for $k \geq 3$}

\subsection{Theorem (Semantic Explosion of $\alpha_f$)}

Let $f$ be a $k$-SAT formula over $n$ Boolean variables, with $m = \delta n$ clauses for some fixed $\delta > 0$. Assume each clause contains $k \geq 3$ literals over disjoint or minimally overlapping sets of variables (i.e., clause independence or bounded local dependency). Then the semantic expansion mapping $\alpha_f : S_{\mathrm{syn}} \rightarrow S_{\mathrm{sem}}$ satisfies the following:

\[
|\alpha_f(S_{\mathrm{syn}})| \in \Omega(c^n) \quad \text{for some constant } c > 1
\]

That is, the size of the semantic space induced by the syntactic clause set grows exponentially in the number of variables.

\subsection{Definitions}

\begin{itemize}
  \item Let $S_{\mathrm{syn}} = \{C_1, \dots, C_m\}$ be the set of CNF clauses.
  \item Each clause $C_i$ is of arity $k$, i.e., $C_i = (\ell_{i1} \vee \dots \vee \ell_{ik})$.
  \item $\alpha_f(C_i) \subseteq \{0,1\}^{V(C_i)}$ is the set of assignments satisfying clause $C_i$.
  \item The total semantic image is defined as: 
  \[
  \alpha_f(S_{\mathrm{syn}}) = \bigcap_{i=1}^m \alpha_f(C_i)
  \]
\end{itemize}

\subsection{Proof}

\textbf{Step 1: Size of Semantic Image per Clause}

Each $k$-literal clause is satisfied by all assignments except one: the all-false configuration. Therefore, for each clause $C_i$:

\[
|\alpha_f(C_i)| = 2^k - 1
\]

\textbf{Step 2: Clause Independence (or Limited Overlap)}

Assume that each clause uses disjoint or minimally overlapping variables. In the worst case, we allow a bounded variable reuse such that dependencies between clauses are negligible. Then the size of the joint semantic space satisfies:

\[
|\alpha_f(S_{\mathrm{syn}})| \geq \prod_{i=1}^m |\alpha_f(C_i)| = (2^k - 1)^m
\]

This becomes tight under full independence, and remains an asymptotic lower bound under weak overlap conditions.

\textbf{Step 3: Relation to Number of Variables}

Assume $m = \delta n$ for some constant $\delta > 0$. Then:

\[
|\alpha_f(S_{\mathrm{syn}})| \in \Omega\left((2^k - 1)^{\delta n}\right) = \Omega(c^n)
\]

where $c = (2^k - 1)^\delta > 1$ for any $k \geq 3$.

\subsection{Conclusion}

The semantic space $\alpha_f(S_{\mathrm{syn}})$ generated by a $k$-SAT formula with $k \geq 3$ grows exponentially with the input size $n$, even under minimal assumptions of clause independence or limited local interaction. Hence, $\alpha_f$ is \textit{explosive} in the sense of the CDF framework.

This formally justifies the classification of $k$-SAT (for $k \geq 3$) as an \textbf{ExpCDF} structure, and highlights the combinatorial explosion inherent in its semantic unfolding.

\section{Formal Proof: Compositionality of Inference in Linear Implicative Structures (2SAT)}

\subsection{Definitions and Assumptions}

Let the following components be defined:

\begin{itemize}
    \item $S_{\text{syn}}$: The set of syntactic clauses in CNF form (e.g., 2SAT).
    \item $S_{\text{sem}}$: The set of semantic implications derived from clauses.
    \item $S_{\text{log}}$: The set of logical outcomes obtained by reasoning (e.g., variable propagation).
\end{itemize}

Define the mappings:

\begin{align*}
\alpha_f &: S_{\text{syn}} \rightarrow S_{\text{sem}} \quad \text{(syntactic to semantic)} \\
\beta_f &: S_{\text{sem}} \rightarrow S_{\text{log}} \quad \text{(semantic to logical inference)} \\
\gamma_f &: S_{\text{syn}} \rightarrow S_{\text{log}} \quad \text{(direct inference from syntax)}
\end{align*}

\textbf{Proposition:}  
For all $C \in S_{\text{syn}}$, if the clause structure is implicative and linear, then:
\[
\gamma_f(C) = \beta_f(\alpha_f(C))
\]

\subsection{Assumptions}

\begin{enumerate}
    \item \textbf{Implication-convertibility:} Any clause $C = (\lnot p \lor q)$ can be rewritten as $p \rightarrow q$ (by definition of $\alpha_f$).
    \item \textbf{Linearity:} $S_{\text{syn}}$ forms a linear implication chain (e.g., $p \rightarrow q \rightarrow r$).
    \item \textbf{Sequential inference:} $\beta_f$ and $\gamma_f$ both perform forward, stepwise reasoning based on implications.
\end{enumerate}

\subsection{Step 1: Base Case (Single Clause)}

Let $C = (\lnot p \lor q) \in S_{\text{syn}}$.

\begin{itemize}
    \item $\alpha_f(C) = p \rightarrow q$
    \item $\beta_f(\alpha_f(C))$ derives $q = \text{true}$ from $p = \text{true}$ (sequential implication).
    \item $\gamma_f(C)$ performs the same propagation directly from the clause.
\end{itemize}

\textbf{Conclusion:} $\gamma_f(C) = \beta_f(\alpha_f(C))$

\subsection{Step 2: Inductive Hypothesis}

Assume that for a chain of $n$ clauses:

\[
C_1, C_2, \dots, C_n \in S_{\text{syn}}
\]

it holds that:

\[
\gamma_f(\{C_1, \dots, C_n\}) = \beta_f(\alpha_f(\{C_1, \dots, C_n\}))
\]

\subsection{Step 3: Inductive Step (n+1 Clauses)}

Add a new clause $C_{n+1} = (\lnot r \lor s)$.

\begin{itemize}
    \item $\alpha_f(C_{n+1}) = r \rightarrow s$
    \item By linearity, $C_n$ and $C_{n+1}$ form a chain (e.g., $p_1 \rightarrow p_2 \rightarrow \cdots \rightarrow r \rightarrow s$)
    \item By the inductive hypothesis:
    \[
    \gamma_f(\{C_1, \dots, C_n\}) = \beta_f(\alpha_f(\{C_1, \dots, C_n\}))
    \]
    \item $\beta_f(\alpha_f(\{C_1, \dots, C_{n+1}\}))$ sequentially derives $s$
    \item $\gamma_f(\{C_1, \dots, C_{n+1}\})$ does the same, via direct propagation
\end{itemize}

\textbf{Conclusion:}
\[
\gamma_f(\{C_1, \dots, C_{n+1}\}) = \beta_f(\alpha_f(\{C_1, \dots, C_{n+1}\}))
\]

\subsection{Step 4: Conclusion by Induction}

Therefore, for all finite linear chains of implicative clauses in $S_{\text{syn}}$, we have:

\[
\forall C \in S_{\text{syn}}, \quad \gamma_f(C) = \beta_f(\alpha_f(C))
\]

\subsection{Appendix: Example in 2SAT}

Let:

\[
C_1 = (\lnot x \lor y), \quad C_2 = (\lnot y \lor z)
\]

Then:

\[
\alpha_f(C_1) = x \rightarrow y, \quad \alpha_f(C_2) = y \rightarrow z
\]

If $x = \text{true}$, then:

\[
\beta_f(\alpha_f(\{C_1, C_2\})) = \{y = \text{true}, z = \text{true}\}
\]
\[
\gamma_f(\{C_1, C_2\}) = \{y = \text{true}, z = \text{true}\}
\]

\textbf{Result:} $\gamma_f = \beta_f \circ \alpha_f$ holds.

\subsection{Summary}

We have formally shown that when syntactic clauses are reducible to implications and form a linear structure, then the direct inference function $\gamma_f$ is functionally equivalent to the composition $\beta_f \circ \alpha_f$.

This justifies the polynomial-time solvability of problems like 2SAT in terms of structural compositionality.

Conversely, in problems like 3SAT, where the clause structure breaks implicative linearity, this identity fails—leading to semantic fragmentation and exponential reasoning complexity.

\section{General Proposition: Non-Compositionality in \textit{k}-SAT for $k \geq 3$}

\subsection{Statement}

\textbf{Proposition.} For any integer $k \geq 3$, there exists a clause $C \in S_{\mathrm{syn}}(f)$ in a $k$-SAT formula such that:
\[
\gamma_f(C) \neq \beta_f(\alpha_f(C))
\]
That is, the direct logical inference from syntax ($\gamma_f$) is not equal to the composition of semantic expansion ($\alpha_f$) followed by logical deduction ($\beta_f$). This demonstrates the failure of compositionality in the CDF mapping for general $k$-SAT.

\subsection{Definitions}

Let $C = (\ell_1 \vee \ell_2 \vee \cdots \vee \ell_k)$ be a clause in CNF form, where $\ell_i$ are literals over distinct Boolean variables, and $k \geq 3$.

We define the three structural mappings of the Construction Defining Functionality (CDF) framework as follows:
\begin{itemize}
    \item $\alpha_f : S_{\mathrm{syn}} \rightarrow S_{\mathrm{sem}}$ maps syntactic clauses to their semantic interpretation (i.e., satisfying assignments).
    \item $\beta_f : S_{\mathrm{sem}} \rightarrow S_{\mathrm{log}}$ maps semantic content to logical inferences.
    \item $\gamma_f : S_{\mathrm{syn}} \rightarrow S_{\mathrm{log}}$ maps syntactic clauses directly to logical consequences.
\end{itemize}

\subsection{Semantic Expansion for $k$-SAT}

Each $k$-ary clause $C = (\ell_1 \vee \cdots \vee \ell_k)$ is satisfied by all truth assignments except the one where all literals are assigned false:
\[
\alpha_f(C) = \{0,1\}^k \setminus \{(0,0,\dots,0)\}
\]
Hence, the size of the semantic image is $|\alpha_f(C)| = 2^k - 1$, which grows exponentially in $k$.

\subsection{Failure of Deductive Determinism}

Unlike in 2-SAT, where clauses can be transformed into binary implications (e.g., $\neg x \vee y \equiv x \rightarrow y$), the clause $C$ for $k \geq 3$ cannot be reduced to a linear implication chain. Consequently:
\begin{itemize}
    \item The mapping $\beta_f(\alpha_f(C))$ cannot deduce the truth value of any individual literal deterministically.
    \item The direct inference $\gamma_f(C)$ requires non-deterministic search, such as branching and backtracking.
\end{itemize}

\subsection{Conclusion}

Therefore, for any $k \geq 3$, there exist syntactic configurations such that:
\[
\gamma_f(C) \neq \beta_f(\alpha_f(C))
\]
This structural non-compositionality arises from:
\begin{itemize}
    \item The inability to reduce high-arity disjunctions to implications.
    \item The exponential growth of the semantic space $\alpha_f(C)$.
    \item The necessity of non-deterministic reasoning in both $\beta_f$ and $\gamma_f$.
\end{itemize}

This proves that compositionality fails in general for $k$-SAT with $k \geq 3$, formally supporting the classification of $k$-SAT under the ExpCDF category in the CDF framework.

\section{Minimal Syntax and the Structural Origin of NP-Hardness} This section synthesizes the formal results on $\alpha_f$-explosion and compositional breakdown, and shows that the syntactic structure of $k$-SAT (for $k \geq 3$) constitutes the minimal generative source of NP-hardness within the CDF framework. 

\subsection{Critical Threshold for Semantic Explosion} As formally established, in $k$-SAT with $k \geq 3$, the syntactic-to-semantic mapping $\alpha_f$ is \textit{explosive}: \begin{quote} \textbf{Theorem.} For any $k \geq 3$, and under mild assumptions of clause independence, the semantic image of $\alpha_f$ satisfies: \[ |\alpha_f(S_{\mathrm{syn}})| \in \Omega(c^n) \quad \text{for some } c > 1 \] \end{quote} Each $k$-literal clause induces a semantic space of size $2^k - 1$, and with $m = \delta n$ clauses, the total space grows as $(2^k - 1)^m = \Omega(c^n)$. This exponential growth emerges from the most compact possible syntactic configuration that permits disjunctive clause composition.

\subsection{Compositionality Breakdown at $k = 3$} In 2SAT, the clause structure admits implicative rewriting and linear propagation. This ensures full compositional coherence: \[ \gamma_f = \beta_f \circ \alpha_f \] In contrast, for any $k \geq 3$, there exist clauses such that: \[ \gamma_f(C) \neq \beta_f(\alpha_f(C)) \] This reflects a structural failure of compositionality. The high-arity disjunctions in 3SAT cannot be reduced to implications, and logical reasoning requires nondeterministic search. Thus, the mapping $\gamma_f$ can no longer be reconstructed from $\alpha_f$ and $\beta_f$.

\subsection{Conclusion: The Generative Core of NP-Hardness}

These results jointly imply that:

\begin{itemize}
  \item Semantic explosion (\textbf{ExpCDF}) arises minimally at $k = 3$;
  \item The transition from 2SAT to 3SAT marks a phase shift from functional stability to combinatorial intractability;
  \item The syntactic features responsible for NP-hardness—namely disjunctive clauses with non-implicative structure—are already fully expressed in 3SAT.
\end{itemize}

Hence, 3SAT is not merely an instance of NP-completeness—it expresses the \textit{minimal syntactic conditions} required to generate NP-hard behavior. All other NP-complete problems derive their intractability via reductions that ultimately encode their structure into the generative syntax of SAT.

From the CDF perspective, the source of NP-hardness is neither algorithmic nor semantic in origin, but fundamentally \textbf{syntactic}.

\subsubsection*{Extentive Examples: Graph Problems under CDF Analysis}

To concretely demonstrate how syntactic structure governs computational tractability, we apply the above principles to classic graph problems:

\paragraph{Perfect Matching (P class):}
Perfect matching can be expressed using binary (2-literal) constraints where each vertex participates in exactly one matched edge. This admits a reduction to 2SAT-like implicative structure. The syntactic composition remains linearly constrained and compositionality holds, i.e., $\gamma_f = \beta_f \circ \alpha_f$ persists. As a result, semantic complexity remains bounded, enabling polynomial-time solutions.

\paragraph{Hamiltonian Cycle (NP-complete):}
In contrast, Hamiltonian Cycle requires selecting a cyclic permutation of vertices such that each appears exactly once. The constraints are inherently high-arity and disjunctive (e.g., “vertex $v$ must follow exactly one of many possible neighbors”), making implicative rewriting infeasible. The compositional mapping breaks down: $\gamma_f \neq \beta_f(\alpha_f)$. This induces semantic explosion analogous to 3SAT, positioning the problem in NP.

\paragraph{Eulerian Path (P class):}
Although historically suspected to be hard, Eulerian Path is solvable in linear time. Its decision condition is purely structural: the number of vertices with odd degree must be 0 or 2. This rule is compositional and global, yet does not require nondeterministic selection among high-arity disjunctions. Thus, the problem remains structurally below the $\alpha_f$-explosion threshold and is tractable.

\subsubsection*{Summary}

These examples confirm that the CDF framework correctly aligns syntactic composition with computational complexity:

\begin{itemize}
  \item Problems in P admit linear, implicative structures (like 2SAT);
  \item NP-complete problems manifest high-arity disjunctive constraints (like 3SAT);
  \item The boundary of tractability aligns with the breakdown of $\gamma_f = \beta_f \circ \alpha_f$.
\end{itemize}

Thus, syntactic generativity—not algorithmic ingenuity—ultimately dictates where computational hardness begins.

\begin{table}[h]
\centering
\begin{tabular}{|l|p{5.5cm}|p{6.5cm}|}
\hline
\textbf{Problem Class} & \textbf{Syntactic Type} & \textbf{Significance} \\
\hline
P Class & Implicative/linear(2SAT-type) & The solution space does not explode exponentially; deterministic computability is guaranteed. \\
\hline
NP-complete & High-arity/disjunctive(3SAT-type) & The problem’s syntax directly induces semantic and logical complexity, causing computational hardness. \\
\hline
\end{tabular}
\end{table}

\section{Functions vs Algorithms: A CDF-Theoretic Distinction}

While functions and algorithms are often treated interchangeably in computer science, the CDF framework reveals a structural distinction between the two. Both can be represented via the same functional notation---e.g., \( f(x) = x + 1 \)---but their internal architecture and operational role are fundamentally different.

\subsection*{Functions as Stable Structural Maps}

Within CDF, a \emph{function} is understood as a coherent structure of three layers linked by the mappings.

If the triangle commutes (\( \gamma_f = \beta_f \circ \alpha_f \)) and all maps are stable, the structure preserves \emph{functionality} in the CDF sense.

\vspace{1em}

\subsection*{Algorithms as Constructive Realizations}

By contrast, an \emph{algorithm} is not defined merely by the input-output relation, but by the constructive derivation \( \gamma_f \)—the actual steps by which output is generated from input.

In this view:

\begin{itemize}
    \item \( \gamma_f \) captures the computational process (e.g., a derivation tree or execution trace).
    \item If \( \gamma_f = \beta_f \circ \alpha_f \), and this computation is finitely realizable, then the function is algorithmically executable.
    \item Otherwise, it is a non-constructive function: defined, but not computable in practice.
\end{itemize}

\begin{table}[h]
\centering
\begin{tabular}{|l|c|c|}
\hline
\textbf{Aspect} & \textbf{Function} & \textbf{Algorithm} \\
\hline
CDF Focus & \( \alpha_f, \beta_f, \gamma_f \) coherence & Constructiveness of \( \gamma_f \) \\
Nature & Declarative (relation) & Procedural (process) \\
Compositionality & Requires triangle to commute & Requires computation to be finite \\
Example & \( f(x) = x + 1 \) (defined) & "Increment x by 1" (realized) \\
Failure Case & \( \alpha_f \) or \( \beta_f \) unstable & \( \gamma_f \) is non-deterministic or explosive \\
\hline
\end{tabular}

\end{table}

\vspace{1em}

\subsection*{When Functionality Breaks: The Case of 3SAT}

Problems such as 3SAT demonstrate the breakdown of functionality:

\begin{itemize}
    \item \( \alpha_f \) becomes \emph{explosive}: small syntactic expressions produce vast semantic spaces.
    \item \( \gamma_f \) no longer equals \( \beta_f \circ \alpha_f \): direct derivation is non-deterministic or backtracking.
    \item The structure loses algorithmic realizability—hence, it ceases to behave as a function.
\end{itemize}

This illustrates how the CDF framework distinguishes between being \emph{functionally defined} and being \emph{computationally realizable}. In this sense, algorithms are not merely implementations of functions, but \textbf{structurally distinct objects} whose coherence depends on the stability of all layers in the CDF diagram.

\section{Structural Reformulation of Computability and Determinism in CDF}

In classical Turing theory, notions such as "computability" and "determinism/non-determinism" are primarily characterized by algorithm termination on inputs and the presence or absence of branching. However, from the perspective of the Compositional Derivation Framework (CDF), these concepts can be understood in a more structural and intrinsic way.

\subsection*{Computability and Constructiveness of \( \gamma_f \)}

Within CDF, a function \( f: X \to Y \) is decomposed into three layers of structure:

\begin{itemize}
  \item \textbf{Syntactic component} \( S_{\text{syn}}(f) \): the syntactic description of the function, e.g., logical formulas or program syntax.
  \item \textbf{Semantic component} \( S_{\text{sem}}(f) \): the semantic interpretation of the syntax, e.g., truth-value spaces or state transitions.
  \item \textbf{Logical component} \( S_{\text{log}}(f) \): the derivation process or computational steps, e.g., derivation trees, proofs, or execution traces.
\end{itemize}

These are connected by three mappings:

\[
\alpha_f: S_{\text{syn}} \to S_{\text{sem}}, \quad
\beta_f: S_{\text{sem}} \to S_{\text{log}}, \quad
\gamma_f: S_{\text{syn}} \to S_{\text{log}}
\]

A function \( f \) is said to be computable if the derivation from syntactic description (input) to logical derivation (output) is \textbf{constructive and achievable in finite steps}, that is:

\[
\gamma_f \text{ is explicitly defined and stably constructible}
\]

\subsection*{Determinism and Commutativity of Mappings}

Furthermore, determinism corresponds to the commutativity of the following diagram:

\[
\gamma_f = \beta_f \circ \alpha_f
\]

which means that:

\begin{itemize}
  \item the syntactic description is stably translated into semantics (\(\alpha_f\)),
  \item semantics are propagated sequentially as logical derivations (\(\beta_f\)),
  \item and consequently, the direct derivation from syntax to logic (\(\gamma_f\)) is consistently constructible.
\end{itemize}

If this compositionality breaks down, semantic explosion or branching inference arises, leading to non-determinism or computational hardness.

\subsection*{Non-determinism and Structural Collapse}

The following is an example of how the breakdown of CDF structure causes non-determinism and computational difficulty:

\begin{itemize}
  \item \textbf{Syntactic failure} (\(\alpha_f\) collapse): clauses cannot be simply reduced to implications; constraints are non-local and entangled.
  \item \textbf{Semantic failure} (\(\beta_f\) instability): semantic space explodes exponentially, preventing local inference.
  \item \textbf{Logical failure} (\(\gamma_f\) non-constructivity): solution search becomes branching and cannot be realized by sequential reasoning alone (e.g., backtracking in SAT solvers).
\end{itemize}

\subsection*{Summary: What Does CDF Visualize?}

Thus, the CDF perspective allows us to view traditional concepts of algorithmic termination and branching not only as procedural or computational phenomena but as properties of the \textbf{structural integrity and mutual coherence} of:

\begin{itemize}
  \item syntactic expressivity (\(\alpha_f\)),
  \item semantic locality and propagatability (\(\beta_f\)),
  \item and logical derivational constructiveness (\(\gamma_f\)).
\end{itemize}

This viewpoint prepares the ground for the next section’s discussion on "CDF-complete functions" versus "Explosive CDF functions," providing a deeper understanding of what it means for a formula or system to behave as a function or as a generator of combinatorial complexity.

\section{CDF-Completeness and the Functionality of Functions}

\subsection*{Definition: CDF-Completeness}

Let \( f: X \rightarrow Y \) be a function, interpreted within the CDF framework as a triple of structured components:
\[
S_{\text{syn}}(f), \quad S_{\text{sem}}(f), \quad S_{\text{log}}(f)
\]
with associated intercomponent mappings:
\[
\alpha_f: S_{\text{syn}} \to S_{\text{sem}}, \quad \beta_f: S_{\text{sem}} \to S_{\text{log}}, \quad \gamma_f: S_{\text{syn}} \to S_{\text{log}}
\]

We say that \( f \) is \textbf{CDF-complete} if the following conditions hold:

\begin{enumerate}
  \item \textbf{Disjunctive Definability:} Each component \( S_{\text{syn}}, S_{\text{sem}}, S_{\text{log}} \) can be inductively or combinatorially generated from primitive symbols or rules.
  
  \item \textbf{Structural Balance:} The three mappings maintain mutual coherence without dominance or collapse, i.e., no one layer trivializes or overwhelms the others.
  
  \item \textbf{Stability of Mappings:} All mappings are stable, in the sense defined below.
\end{enumerate}

\subsection*{Definition: Stability and Explosiveness}

Let \( \varphi: A \to B \) be a structural map with size or complexity measures \( |\cdot|: A \cup B \to \mathbb{R}_{\geq 0} \). Then:

\begin{itemize}
  \item \textbf{Stability:} \( \varphi \) is \textit{stable} if there exist constants \( C > 0 \), \( k \in \mathbb{N} \) such that:
  \[
  \forall a \in \text{dom}(\varphi), \quad |\varphi(a)| \leq C \cdot |a|^k
  \]
  
  \item \textbf{Explosiveness:} \( \varphi \) is \textit{explosive} if there exists a polynomially bounded sequence \( (a_n) \subset A \) such that:
  \[
  \limsup_{n \to \infty} \frac{\log |\varphi(a_n)|}{\log n} = +\infty
  \]
\end{itemize}

\subsection*{ComCDF vs ExpCDF: Functional vs Explosive Behavior}

From this perspective:

\begin{itemize}
  \item \textbf{ComCDF} (e.g., 2SAT) satisfies all criteria of CDF-completeness. The structure is function-like, stable, and compositional. It preserves the identity of a function as a coherent input-output mechanism.

  \item \textbf{ExpCDF} (e.g., 3SAT and general \( k \)-SAT with \( k \geq 3 \)) violates stability in at least the \( \alpha_f \) mapping, due to semantic explosion. This breaks the internal balance and causes the structure to lose function-like behavior.
\end{itemize}

Thus, we may interpret CDF-completeness as a \textit{semantic criterion for functionhood}.  
A formula or system that fails this criterion, like 3SAT, ceases to behave as a function in any meaningful sense—it becomes a \textbf{symbolic generator of combinatorial complexity}, rather than a mechanism for constructive transformation.

\section{Redefining NP-Hardness from a CDF Perspective}

\subsection*{From Algorithmic Complexity to Structural Instability}

Traditionally, NP-hardness is understood in terms of algorithmic complexity: a problem is NP-hard if it is at least as hard as the hardest problems in NP, typically evidenced by polynomial-time reductions.

However, this external perspective fails to capture the \textit{internal structural origins} of computational hardness.

The Construction Defining Functionality (CDF) framework offers a new framework: NP-hardness can be understood as a \textbf{breakdown of compositional functionality} across three structural layers.

\subsection*{The Role of \( \alpha_f \): Semantic Explosion as the Root of Hardness}

In compositional systems (ComCDF), the mapping \( \alpha_f \) is stable: the semantic space grows in a polynomially bounded manner with respect to the syntactic input. This preserves functional coherence across layers, and logical reasoning remains tractable.

In explosive systems (ExpCDF), such as 3SAT or general \( k \)-SAT for \( k \geq 3 \), the mapping \( \alpha_f \) becomes \textbf{explosive}. The number of semantic configurations generated from a compact syntactic clause grows super-polynomially, leading to:

\begin{itemize}
  \item \textbf{Non-local and fragmented semantic spaces}
  \item \textbf{Non-deterministic and branching logical inference}
  \item \textbf{Breakdown of function-like behavior}
\end{itemize}

Thus, the system loses its CDF-completeness—it no longer behaves like a coherent function from input to output.

\subsection*{NP-Hardness as Loss of Functional Identity}

We propose the following structural reinterpretation:

\begin{quote}
\textit{A problem is NP-hard not merely because it lacks a known efficient algorithm, but because its internal structure fails to satisfy CDF-completeness.}

\textit{The instability (explosiveness) of the syntactic-to-semantic map \( \alpha_f \) destroys the functional identity of the system, rendering it computationally intractable.}
\end{quote}

This explains why SAT is not just a representative of NP-hardness, but a \textbf{generator} of computational hardness: it encodes the very loss of functionhood at the structural level.

\subsection*{Diagrammatic Summary}

\begin{center}
\begin{tabular}{|l|c|c|}
\hline
\textbf{Property} & \textbf{ComCDF (e.g., 2SAT)} & \textbf{ExpCDF (e.g., 3SAT)} \\
\hline
Mapping \( \alpha_f \) & Stable & Explosive \\
Semantic Space & Local, Constructive & Fragmented, Explosive \\
Logic Inference & Deterministic, Linear & Non-deterministic, Backtracking \\
Functionality & Preserved & Collapsed \\
NP-Hardness & Absent & Emergent \\
\hline
\end{tabular}
\end{center}

\subsection*{Toward a Structural Theory of Complexity}

This perspective opens the door to a structural, rather than algorithmic, theory of complexity:

\begin{itemize}
  \item Complexity arises not from size alone, but from the \textbf{loss of internal compositionality}.
  \item NP-hardness reflects the \textbf{inability to preserve functional identity} across syntactic, semantic, and logical layers.
  \item The CDF framework can serve as a foundation for redefining complexity classes in terms of structural properties.
\end{itemize}

\section{Structural Origins of Complexity and the Fragility of Problem Identity}

\paragraph{Syntactic generativity as the source of semantic explosion}

Within the CDF framework, the semantic intractability of problems such as SAT is not an incidental phenomenon, but rather a structurally deterministic outcome of their syntactic formulation.

Clause-based constructions—particularly in CNF with disjunctive connections and non-local literal interactions—impose structural configurations in \( S_{\text{syn}} \) that inherently lead to highly fragmented and non-constructive semantic spaces in \( S_{\text{sem}} \). This process is governed by the map:

\[
\alpha_f : S_{\text{syn}} \rightarrow S_{\text{sem}}
\]

When \( \alpha_f \) exhibits exponential or super-polynomial growth, even compact syntactic expressions give rise to combinatorially explosive semantic configurations. Thus, the complexity observed at the level of meaning is not introduced externally, but rather generated internally from the syntactic architecture itself.

\paragraph{Controlling complexity through syntactic constraint}

From this perspective, any attempt to reduce semantic complexity must necessarily act at the syntactic level—specifically, by constraining the generative rules that define \( S_{\text{syn}} \). That is, one must suppress the explosiveness of \( \alpha_f \).

Common strategies in SAT variants illustrate this principle:

\begin{itemize}
  \item Limiting clause width (e.g., 3SAT to 2SAT),
  \item Imposing structural conditions (e.g., Horn clauses, bounded treewidth),
  \item Restricting variable interactions or clause connectivity.
\end{itemize}

Such interventions can yield tractable sub-classes by enforcing a stable, compositional mapping \( \alpha_f \), thereby restoring functional coherence across the CDF layers.

\paragraph{Complexity reduction and the erosion of problem identity}

However, these syntactic constraints do not merely simplify the problem—they transform it. The generative structure of the original problem, which defines its identity within a complexity class, is no longer preserved.

Hence, we assert the following:

\begin{quote}
To render an NP-complete problem tractable is to alter its syntactic construction in such a way that the explosive growth of semantic configurations is suppressed. This entails a redefinition of the problem’s structural identity.
\end{quote}

In this sense, the pursuit of polynomial-time solutions to NP-complete problems is not simply an algorithmic endeavor, but a structural one. It requires modifying the very rules by which the problem is defined—rules that generate its internal complexity.

Accordingly, the CDF framework reveals a fundamental trade-off: the structural source of complexity is identical to the structural source of the problem’s identity. Efforts to control the former may unavoidably compromise the latter.

\paragraph{Final Remark.}
In this paper, we have shown that the compositional collapse observed in NP-complete problems such as SAT 
is not a consequence of algorithmic lack , but of syntactic inevitability.

Through the CDF framework, the exponential growth of the syntax-to-semantics mapping \( \alpha_f \) 
reveals that semantic explosion is not provided from the outside but structurally generated. 
This structural generation breaks the functional coherence \( \gamma_f = \beta_f \circ \alpha_f \), 
rendering such problems \emph{functionally unstable}.

Therefore, if one were to devise a polynomial-time algorithm for an NP-complete problem, 
it would require redefining the very syntactic rules that generate its complexity—altering the problem’s identity. 

In this sense, \( \mathrm{P} \neq \mathrm{NP} \) is not merely a complexity-theoretic conjecture, 
but a reflection of structural incompatibility between generative syntax and tractable computation.

\nocite{*}
\bibliographystyle{plain}
\bibliography{references}

\end{document}